\documentclass[aps,twocolumn,prd,showpacs,nofootinbib]{revtex4}
\usepackage{amsmath}
\usepackage{graphicx}
\usepackage{subfigure}
\usepackage{dcolumn}
\usepackage{bm}
\usepackage{amssymb}
\usepackage{latexsym}

\def\setC{\mathbb{C}}

\def\setR{\mathbb{R}}

\newcommand{\gsim}{\gtrsim}
\newcommand{\lsim}{\lesssim}
\newcommand{\ie}{\textsl{i.e.~}}

\newcommand{\mP}{m_{_{\mathrm Pl}}}
\newcommand{\GReCO}{${\cal G}\setR\varepsilon\setC{\cal O}$}

\def\spose#1{\hbox to 0pt{#1\hss}}
\def\lta{\mathrel{\spose{\lower 3pt\hbox{$\mathchar"218$}}
     \raise 2.0pt\hbox{$\mathchar"13C$}}}
\def\gta{\mathrel{\spose{\lower 3pt\hbox{$\mathchar"218$}}
     \raise 2.0pt\hbox{$\mathchar"13E$}}}

\newcommand{\de}[2]{\kern - #1 em \mathrm{d} #2}

\begin{document}

\title{Birefringent Gravitational Waves and the Consistency Check of
Inflation}

\author{Stephon Alexander}
\email{stephon@itp.stanford.edu}
\affiliation{Stanford Linear Accelerator Center, Stanford University, 
2575 Sand Hill Rd., Menlo Park, CA 94025 USA}

\author{J\'er\^ome Martin} 
\email{jmartin@iap.fr}
\affiliation{Institut d'Astrophysique de
Paris, \GReCO, FRE 2435-CNRS, 98bis boulevard Arago, 75014 Paris,
France}

\date{\today}

\begin{abstract}
In this work we show that the gravitational Chern-Simons term, aside
from being a key ingredient in inflationary baryogenesis, modifies
super-horizon gravitational waves produced during inflation. We
compute the super-Hubble gravitational power spectrum in the slow-roll
approximation and show that its overall amplitude is modified while
its spectral index remains unchanged (at leading order in the
slow-roll parameters). Then, we calculate the correction to the tensor
to scalar ratio, $T/S$.  We find a correction of $T/S$ which is
dependent on $\cal{N}$ (more precisely quadratic in ${\cal N}$), the
parameter characterizing the amplitude of the Chern-Simons terms. In a
stringy embedding of the leptogenesis mechanism, $\cal{N}$ is the
ratio between the Planck scale and the fundamental string scale. Thus,
in principle, we provide a direct probe of leptogenesis due to stringy
dynamics in the Cosmic Microwave Background (CMB). However, we
demonstrate that the corresponding correction of $T/S$ is in fact very
small and not observable in the regime where our calculations are
valid. To obtain a sizable effect, we argue that a non-linear
calculation is necessary.
\end{abstract}

\pacs{98.80.Cq, 98.70.Vc}
\maketitle

\section{Introduction}

Cosmic baryogenesis stands as one of the unresolved problems of
particle cosmology. Most models address baryogenesis after the
inflationary epoch. Recently the authors of Ref.~\cite{APS}
demonstrated that the baryon asymmetry can be generated during
inflation from gravity waves. In this model the lepton number was
generated by a quantum expectation value of the Chern-Simons density
from Ultra-Violet (UV), birefringent gravitational waves during the
inflationary epoch. In a subsequent paper the authors showed that this
model can be embedded in string theory, in a model independent manner,
through the Green-Schwarz mechanism~\cite{Gates}. In the stringy
embedding there was a huge enhancement of the lepton asymmetry due to
a hierarchy in the fundamental string scale and the four dimensional
Planck scale.

\par
  
Other investigators have searched for parity violation in the Cosmic
Microwave Background (CMB) which ultimately leads to a birefringence
in gravitational waves ~\cite{Lue,Pogosian,Balaji}. In this note, we
shall study the super-horizon power spectrum and tensor to scalar
ratio of scalar and tensor birefringent perturbations produced during
inflation. Specifically, we study the spectrum of super-horizon
gravity waves whose UV counterparts were responsible for
leptogenesis. Is it possible to see a signature of the leptogenesis
mechanism in the super-horizon power spectrum?  To address this
question we shall derive the tensor to scalar ratio and show that it
contains a direct signature of the leptogenesis mechanism which
occurred in the UV. Furthermore we show that the scalar to tensor
ratio contains the string scale in a model independent way and is in
an observable window to this physics.

\par

The paper is organized as follows. In section~II we derive the
equations for the gravitational waves in the presence of the
Chern-Simons term. In section~III we provide the exact solutions at
various scales and derive the power spectrum as well as the corrected
tensor to scalar ratio. In section~IV we relate this modified tensor
to scalar ratio to the stringy embedding of gravitational leptogenesis
and we conclude with some open issues concerning a consistent
quantization and further directions.

\section{Basics Equations}

The starting point of inflationary leptogenesis is the
Einstein-Hilbert action coupled to the gravitational Chern-Simons
term, which is necessarily present in string theory. This last term
can be written as
\begin{equation}
\label{CSaction}
S_{_{\rm CS}} ={1\over 8\kappa}\int {\rm d}^{4}x f(\phi)R\wedge R \, ,
\end{equation}
where $\kappa \equiv 8\pi /\mP^2$, $\mP$ being the Planck mass. We
proceed to linearize the Einstein-Hilbert action with the Chern-Simons
term in a Friedmann-Lema\^{\i}tre-Robertson-Walker (FLRW) background
in the presence of tensor perturbations (\ie in presence of
gravitational waves). The corresponding metric tensor takes the form
(assuming that the space-like sections are flat)
\begin{eqnarray}
\label{ds2}
{\rm d}s^2 &=& a^2(\eta )\left[- {\rm d}\eta ^2 +\left(\delta _{ij}
+h_{ij}\right){\rm d}x^i{\rm d}x^j\right] \, ,
\end{eqnarray}
with $h_{ij}$, being a transverse and traceless tensor, i.e. $\delta
^{ij}h_{ij}=0$, $\partial ^jh_{ij}=0$ and $a(\eta )$, the FLRW scale
factor, being a function of the conformal time $\eta $. Due to the
symmetries of the FLRW metric, the inflaton field $\phi $ in
Eq.~(\ref{CSaction}) is also a function of the conformal time only.

\par

Expanding the action up to second order in the gravitational waves
tensor $h_{ij}$ (which is necessary in order to obtain first order
equations of motion), after lengthy but straightforward calculations,
one obtains the following expression
\begin{widetext}
\begin{eqnarray}
\label{action}
{}^{(2)}S_{_{\rm GW}} &=& \frac{1}{8\kappa }\int {\rm
d}^4x\biggl\{a^2(\eta )
\biggl[\left(h^i{}_j\right)'\left(h^j{}_i\right)' -\left(\partial
_kh^i{}_j\right)\left(\partial ^kh^j{}_i\right)\biggr] -f'\epsilon
^{ijk} \biggl[\left(h^q{}_i\right)'\left(\partial _jh_{kq}\right)'
-\left(\partial ^rh^q{}_i\right)\partial _j\partial
_rh_{kq}\biggr]\biggr\}\, ,
\end{eqnarray}
where a prime stands for a derivative with respect to conformal time
and $\epsilon ^{ijk}\equiv \epsilon ^{0ijk}$, $\epsilon ^{\mu \nu \tau
\sigma }$ being the totally antisymmetric tensor.  In the above
expression, one recognizes the standard (\ie Einstein-Hilbert)
expression of the perturbed action (first term between squared
brackets) while the term proportional to $f'$ represents the
correction coming from the Chern-Simons contribution.  Varying this
action with respect to the gravitational waves tensor, one obtains the
first order equation of motion which reads
\begin{eqnarray}
\left(h^j{}_i\right)''+2\frac{a'}{a}\left(h^j{}_i\right)'
-\partial _k\partial ^kh^j{}_i
+\frac{1}{a^2}\epsilon^{pjk}\biggl[f''\left(\partial _ph_{ki}\right)'
+f'\left(\partial _ph_{ki}\right)''-f'\partial _p\partial ^r\partial _r
h_{ki}\biggr] &=& 0 \, ,
\end{eqnarray}
\end{widetext}
Next, following Ref.~\cite{hwang}, we define the tensor $D_{ij}$ by
the following equation
\begin{equation}
D_{ij}\equiv h_{ij}''+2\frac{a'}{a}h_{ij}'-\partial _k\partial ^kh_{ij}\, ,
\end{equation}
and, then, the equation of motion takes the form
\begin{equation}
D^j{}_i+\frac{1}{a^2}\epsilon ^{pjk}\left[\left(f''-2{\cal H}f'\right)
\partial _ph_{ki}'+f'\partial _pD_{ki}\right]=0 \, ,
\end{equation}
where we have defined ${\cal H}\equiv a'/a$. This equation is similar
to Eqs.~(11) and (12) of Ref.~\cite{hwang}, except that we have
written the equation in terms of the conformal time rather than in
terms of the cosmic time.

\par

The next step consists in going to the Fourier space. For this
purpose, we write the metric tensor as
\begin{equation}
h_{ij}\left(\eta ,{\mathbf x}\right)=\frac{1}{\left(2\pi \right)^{3/2}}
\int {\rm d}{\mathbf k}\sum _{s=1}^2p_{ij}^s\left({\mathbf k}\right)
h_{\mathbf k}^s\left(\eta \right){\rm e}^{i{\mathbf k}\cdot {\mathbf x}}\, .
\end{equation}
In the above expression, $p_{ij}^s\left({\mathbf k}\right)$ is the
linear polarization tensor ($s=1,2$ corresponds to $s=+,\times
$). Concretely, if the wave-vector is written in polar coordinates as
${\mathbf k}/k=\left(\sin \theta \cos \varphi, \sin \theta \sin
\varphi , \cos \theta \right)$, then two vectors perpendicular to
${\bf k}$ are given by $ {\mathbf e}_1 =\left(\sin \varphi, -\cos
\varphi, 0\right)$ for the first vector and $ {\mathbf e}_2
=\left(\cos \theta \cos \varphi, \cos \theta \sin \varphi, -\sin
\theta \right)$ for the second vector but only if $\theta <\pi /2$. If
$\theta >\pi /2$, {\it i.e.}  if the wave-vector points to the bottom,
the expression of the second vector should in fact read $ {\mathbf
e}_2=-\left(\cos \theta \cos \varphi, \cos \theta \sin \varphi, -\sin
\theta \right)$. It is also interesting to mention how these
quantities transform under the change ${\bf k} \rightarrow -{\mathbf
k}$. It is easy to see that this corresponds to the transformation
$\left(\theta \, ,\varphi \right) \rightarrow \left(\pi -\theta \,
,\varphi +\pi \right)$. Then, we have ${\bf e}_1\rightarrow -{\bf
e}_1$ and ${\mathbf e}_2\rightarrow -{\bf e}_2$.  Finally, the
polarization tensor can be written as
\begin{eqnarray}
p_{ij}^1 &=& \left({\mathbf e}_1\right)_i\left({\mathbf e}_1\right)_j
-\left({\mathbf e}_2\right)_i\left({\mathbf e}_2\right)_j\, ,
\\
p_{ij}^2 &=& \left({\mathbf e}_1\right)_i\left({\mathbf e}_2\right)_j
+\left({\mathbf e}_1\right)_j\left({\mathbf e}_2\right)_i\, .
\end{eqnarray}
Due to the properties of the vectors ${\bf e}_1$ and ${\bf e}_2$
established above, it is easy to check that $p_{ij}^s\left(-{\mathbf
k}\right)=p_{ij}^s\left({\mathbf k}\right)$ and
$p_{ij}^s\left({\mathbf k}\right)p^{ij}{}^{s'}\left({\mathbf k}\right)
=2\delta ^{ss'}$. Using these properties and the fact that $h_{ij}$ is
real, $h_{ij}=h_{ij}^*$, one can also establish that
\begin{equation}
\left(h_{\mathbf k}^s\right)^*=h_{-\mathbf k}^s\, ,\quad s=+\, ,\times \, .
\end{equation}
The next step consists in defining two other states of polarization,
the so-called right and left polarization states. The corresponding
polarization tensors are given by
\begin{eqnarray}
p_{ij}^{\rm R} &\equiv & \frac{1}{\sqrt{2}}
\left(p_{ij}^1+ip_{ij}^2\right)\, ,
\\ 
p_{ij}^{\rm L} &\equiv & 
\frac{1}{\sqrt{2}}\left(p_{ij}^1-ip_{ij}^2\right)=
\left(p_{ij}^{\rm R}\right)^*\, ,
\end{eqnarray}
{F}rom the above expressions, using the properties of the linear
polarization tensors, one can show that
\begin{eqnarray}
p_{ij}^{\rm R}\left({\bf k}\right)p^{ij}{}^{\rm R}\left({\bf k}\right)
&=& p_{ij}^{\rm L}\left({\bf k}\right)p^{ij}{}^{\rm L}\left({\bf
k}\right)=0\, , 
\\ 
p_{ij}^{\rm R}\left({\bf k}\right)p^{ij}{}^{\rm
L}\left({\bf k}\right)&=&2\, .
\end{eqnarray}
These expressions are of course valid only if the polarization tensors
are evaluated for the same wave-number. We also have
$p_{ij}^s\left({\bf k}\right) = p_{ij}^s\left(-{\bf k}\right)$ with
$s={\rm R}, {\rm L}$. Then, using the expression of the vectors
${\mathbf e}_1$ and ${\mathbf e}_2$, it is easy to show that
\begin{equation}
\label{eigenpola}
\frac{k_p}{k}\epsilon ^{mpj}p_{ij}^s
=\mp i\lambda ^s\left(p^m{}_i\right)^s\, ,\quad s={\rm R}, {\rm L}\, .
\end{equation}
where $\lambda ^{\rm R}=+1$ and $\lambda ^{\rm L}=-1$ and where the
upper sign refers to $\theta <\pi /2$ while the lower one refers to
$\theta >\pi /2$.

\par

We are now in a position where one can re-write the gravitational
waves tensor in terms of the left and right polarization states. This
gives
\begin{equation}
\label{gwpolaLR}
h_{ij}\left(\eta ,{\mathbf x}\right)=\frac{1}{\left(2\pi \right)^{3/2}}
\int {\rm d}{\mathbf k}\sum _{s={\rm R},{\rm L}}p_{ij}^s
\left({\mathbf k}\right)
h_{\mathbf k}^s\left(\eta \right){\rm e}^{i{\mathbf k}\cdot {\mathbf x}}\, ,
\end{equation}
where we have introduced the definitions
\begin{eqnarray}
\label{link1}
h_{\bf k}^{\rm R} &\equiv & \frac{1}{\sqrt{2}}
\left(h_{\bf k}^1-ih_{\bf k}^2\right)\, ,
\quad  
h_{\bf k}^{\rm L} \equiv  
\frac{1}{\sqrt{2}}\left(h_{\bf k}^1+ih_{\bf k}^2\right)\, .
\end{eqnarray}
Let us notice that it is straightforward to demonstrate that
$h_{-{\mathbf k}}^{\rm R}=\left(h_{\mathbf k}^{\rm L}\right)^*$ and
$h_{-{\mathbf k}}^{\rm L}=\left(h_{\mathbf k}^{\rm R}\right)^*$.

\par

The next step consists in introducing the new expansion of the
gravitational waves tensor given by Eq.~(\ref{gwpolaLR}) into the
equation of motion and in using Eq.~(\ref{eigenpola}) to arrive at a
new form of the equation of motion. One obtains
\begin{widetext}
\begin{equation}
\label{eq:motionfourier}
\left(1-\lambda ^sk\frac{f'}{a^2}\right)\left(h_{\bf k}^s\right)''
+\left(2{\cal H}-\lambda ^sk\frac{f''}{a^2}\right)\left(h_{\bf k}^s\right)'
+\left(1-\lambda ^sk\frac{f'}{a^2}\right)k^2h_{\bf k}^s=0\, , \quad 
s={\rm R}\, ,{\rm L}\, .
\end{equation}
\end{widetext}
Finally, the last step consists in introducing the quantity $z_s$
defined by
\begin{equation}
\label{defz}
z_s \left(\eta, {\mathbf k}\right)\equiv a\left (\eta \right)
\sqrt{1-\lambda ^sk\frac{f'}{a^2}}\,
\end{equation}
and the new amplitude $\mu _{\mathbf k}^s(\eta )$ defined by $\mu
_{\mathbf k}^s\equiv z_s h_{\mathbf k}^s$. Then, the equation of motion
for $\mu _{\mathbf k}^s$ has the traditional form of the equation of
motion for a parametric oscillator, namely
\begin{equation}
\label{eq:motionmu}
\left(\mu _{\mathbf k}^s\right)''
+\left(k^2-\frac{z_s''}{z_s}\right)\mu _{\mathbf k}^s=0\, .
\end{equation}
The effective potential $z_s''/z_s$ depends on time, on polarization
(birefringence) but also on the wave-number which is an important
difference with respect to the standard case where the effective
potential depends on conformal time only. This equation has been
derived for the first time in Ref.~\cite{hwang}; see Eq.~(15) of that
paper. However, in Ref.~\cite{hwang}, it is also assumed that the
effective potential takes the form $z_s''/z_s=n_s/\eta ^2$ where $n_s$
is a constant. In particular, one notices that, with this ansatz, the
scale dependence of the effective potential has disappeared. This
permits to find simple solutions in terms of Bessel
functions. However, we will see that, in the present context, the
effective potential is different and more complicated.

\section{Gravitational Waves Power Spectrum in the slow-roll approximation}

\subsection{The Effective Potential}

Let us now calculate the effective potential explicitly. Using the
formulas established previously, one obtains that the exact expression
of the potential can be written as
\begin{eqnarray}
\label{poteff}
\frac{z_s''}{z_s} &=& \frac{a''}{a}-{\cal H}\lambda ^sk
\frac{\left(f'/a^2\right)'}{1-\lambda ^sk\left(f'/a^2\right)}
-\frac{\lambda ^sk}{2}\frac{\left(f'/a^2\right)''}{1-\lambda
^sk\left(f'/a^2\right)} 
\nonumber \\
& & -\frac{1}{4}\left(\lambda ^sk\right)^2
\frac{\left[\left(f'/a^2\right)'\right]^2}{\left[1-\lambda
^sk\left(f'/a^2\right)\right]^2}\, .
\end{eqnarray}
To go further, we need to postulate the function $f$. Following
Ref.~\cite{APS}, we choose
\begin{equation}
f=\frac{{\cal N}}{16\pi ^2M_{_{\rm Pl}}^2}\frac{ \phi }{M_{_{\rm Pl}}}\, .
\end{equation}
where $M_{_{\rm Pl}}\equiv m_{_{\rm Pl}}/\sqrt{8\pi }$ is the reduced
Planck mass and ${\cal N}$ a number that we discuss in more details in
the last section and that can be related to the string scale. With
this definition $f/a^2$ is dimensionless, as it should, if the scale
factor has the dimension of a length (which is our convention). In
terms of the slow-roll parameters $\epsilon \equiv -\dot{H}/H^2$,
$\delta \equiv -\ddot{\phi }/(H\dot{\phi })$ and $\xi \equiv
(\dot{\epsilon }-\dot{\delta })/H$ (a dot means a derivative with
respect to cosmic time), we have at leading order in the slow-roll
parameters, see also Ref.~\cite{MS2}
\begin{eqnarray}
a(\eta ) &=& \ell _0(-\eta )^{-1-\epsilon }\, ,
\quad 
\phi ' \simeq -M_{_{\rm Pl}}
{\cal H}\sqrt{2\epsilon}\, .
\end{eqnarray}
{F}rom this expression, one deduces that (at leading order in the
slow-roll parameters)
\begin{equation} \frac{f'}{a^2}=-\frac{{\cal N}}{16\pi
^2M_{_{\rm Pl}}^2} \frac{\cal H}{a^2}\sqrt{2\epsilon}\simeq
\frac{{\cal N}}{16\pi ^2}\left(\frac{H_{_{\rm
inf}}}{M_{_{\rm Pl}}}\right)^2 \sqrt{2\epsilon }\eta
\end{equation}
because ${\cal H}\simeq -(1+\epsilon)/\eta $. From that expression,
one arrives at the two following formulas which are useful in order to
calculate the effective potential
\begin{eqnarray}
\left(\frac{f '}{a^2}\right)' &=& 
\frac{{\cal N}}{16\pi ^2M_{_{\rm Pl}}^2}
\frac{{\cal H}^2}{a^2}\left(1+\delta \right)
\sqrt{2\epsilon}\, 
\\
&\simeq & \frac{{\cal N}}{16\pi ^2}
\left(\frac{H_{_{\rm inf}}}{M_{_{\rm Pl}}}\right)^2
\sqrt{2\epsilon}+{\cal O}\left(\epsilon ^{3/2}\right)\, ,
\end{eqnarray}
and
\begin{eqnarray}
\left(\frac{f '}{a^2}\right)'' &=& 
\frac{{\cal N}}{16\pi ^2M_{_{\rm Pl}}^2}
\frac{{\cal H}^3}{a^2}\left(-\epsilon -\delta -3\epsilon\delta 
+2\epsilon ^2-\delta ^2-\xi\right)\sqrt{2\epsilon}\, 
\\
&\simeq & \frac{{\cal N}}{16\pi ^2}\left(\frac{H_{_{\rm
inf}}}{M_{_{\rm Pl}}}\right)^2\frac{1}{\eta }\left(\epsilon +\delta\right)
\sqrt{2\epsilon}+{\cal O}\left(\epsilon ^{5/2}\right)\, ,
\end{eqnarray}
Inserting the above equations into the formula giving the potential,
namely Eq.~(\ref{poteff}), one obtains
\begin{widetext}
\begin{eqnarray}
\label{potexplicit}
\frac{z_s''}{z_s} &\simeq & \frac{2+3\epsilon }{\eta ^2} -\lambda
^s\frac{k}{\vert \eta \vert}\frac{{\cal N}}{16\pi ^2}
\left(\frac{H_{_{\rm inf}}}{M_{_{\rm Pl}}}\right)^2 \sqrt{2\epsilon }
\left[1-\lambda
^s\frac{{\cal N}}{16\pi ^2}
\left(\frac{H_{_{\rm inf}}}{M_{_{\rm Pl}}}\right)^2 \sqrt{2\epsilon }
\times (k\eta )\right]^{-1}
\nonumber \\
& & -\frac{k^2}{4}\frac{{\cal N}^2}{256\pi ^4}\left(\frac{H_{_{\rm
inf}}}{M_{_{\rm Pl}}}\right)^4(2\epsilon) 
\left[1-\lambda
^s\frac{{\cal N}}{16\pi ^2}
\left(\frac{H_{_{\rm inf}}}{M_{_{\rm Pl}}}\right)^2 \sqrt{2\epsilon }
\times (k\eta )\right]^{-2}\, ,
\end{eqnarray} 
\end{widetext}
where we have ignored sub-dominant term in the slow-roll
parameters. It is important to notice that, in order to obtain the
above equation, we have never expanded a term like $1-\lambda
^sk\left(f'/a^2\right)$ in the slow-roll parameters. We notice the
presence of $k$ in the numerator of the second term. This is in full
agreement with Ref.~\cite{MS} where it has been noticed that a term
like $1/\vert \eta \vert $ in the effective potential necessarily
implies a new characteristic scale. Here, the characterized scale
defined in Ref.~\cite{MS} could be written as (at this level, the two
situations are not yet totally equivalent because the above effective
potential is not exactly similar to the potential studied in
Ref.~\cite{MS}. This will be the case below.)  Now let
\begin{equation}
k_{_{\rm C}}\equiv k\frac{{\cal N}}{32\pi ^2} \left(\frac{H_{_{\rm
inf}}}{M_{_{\rm Pl}}}\right)^2 \sqrt{2\epsilon }=k\frac{\Theta }{16}\, ,
\end{equation}
where we have defined $\Theta $ by the following relation. (see
Eq.~(13) in Ref.~\cite{APS}):
\begin{equation}
\label{deftheta}
\Theta \equiv \frac{{\cal N}}{2\pi ^2}
\left(\frac{H}{M_{_{\rm Pl}}}\right)^2\sqrt{2\epsilon}\, .
\end{equation}
In the present context, somehow, the characteristic scale $k_{_{\rm
C}}$ ``depends on the scale'' (\ie on $k$). However, we see that the
large-scale limit, as defined in Ref.~\cite{MS} \ie $k\ll k_{_{\rm
C}}$, corresponds in the present context to the condition $\Theta
/16\gg 1$. The only way to satisfy this condition is to have a large
${\cal N}$ which could compensate the smallness of $H/M_{_{\rm Pl}}$
and of the slow-roll parameter.

\par

\begin{figure*}[t]
  \includegraphics[width=.90\textwidth,height=.50\textwidth]{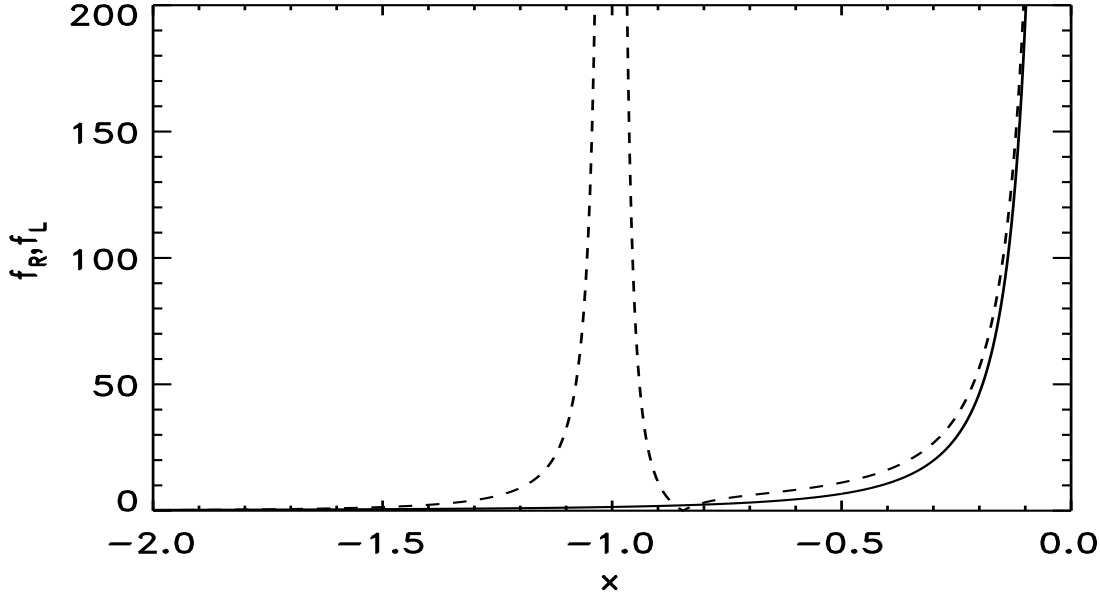} 
\caption{Effective potential for the two states of polarization (solid
line for the right polarization state and dashed line for the left
polarization state). At $x=-1$ or $\eta =-8/(k\Theta )$, the effective
potential $f_{_{\rm L}}(x)$ blows up. For $x>-1$, the slight
difference between $f_{_{\rm L}}(x)$ and $f_{_{\rm R}}(x)$
mathematically originates from the term $\lambda ^s/[x(1-\lambda
^sx)]$ in $z_s''/z_s$ and, physically, from the phenomenon of
birefringence. As $x\rightarrow 0$, the standard term
$(2+3\epsilon)/x^2$ dominates. Since this term does not depend on the
polarization state, one has $f{_{\rm R}}(x)\rightarrow f{_{\rm
L}}(x)$\label{pot}.}
\end{figure*}

For convenience, we now introduce the variable $x$ defined by $x\equiv
\Theta k\eta /8<0$. Then, the equation of motion takes the form
\begin{equation}
\frac{{\rm d}^2\mu }{{\rm d}x^2}+\left[\frac{64}{\Theta ^2}
-f_{_{\rm R,L}}(x)\right]\mu =0\, ,
\end{equation} 
with 
\begin{eqnarray}
f_{_{\rm R}}(x) &=& \frac{2+3\epsilon }{x^2}+\frac{1}{x(1-x)}-\frac14 
\frac{1}{(1-x)^2}\, ,
\\
f_{_{\rm L}}(x) &=& \frac{2+3\epsilon }{x^2}-\frac{1}{x(1+x)}-\frac14 
\frac{1}{(1+x)^2}\, .
\end{eqnarray}
The functions $f_{_{\rm R}}$ and $f_{_{\rm L}}$ are represented in
Fig.~\ref{pot}. From this figure, the different behavior of the two
states of polarization is apparent. The L mode (dashed line)
undergoes a ``kick'' at $x=-1$ where the effective potential blows up.
At the same point the potential of the R mode is perfectly regular (solid
line). Therefore, we expect the R mode function to propagate
smoothly through $x=-1$ while the behavior of the L mode function
can be more problematic. We now turn to this question in more details.

\subsection{Solutions to the Mode Equation in the Vicinity of 
the Divergence}

Let now us study the equation of motion for the left mode in the
vicinity of $x\simeq -1$. It is easy to check that a very good
approximation of the potential is
\begin{eqnarray}
\label{approxpotdiv}
f_{_{\rm L}}(x) &\simeq & \frac{1}{(1+x)}-\frac14 
\frac{1}{(1+x)^2}\, .
\end{eqnarray}
In fact the approximation is good even far form $x\simeq -1$ provided
$x<-1$ since, on small scales, \ie in the limit $k\eta \rightarrow
+\infty$, we have $z_s''/z_s \rightarrow 0$. In the limit, the
solution can be written as
\begin{equation}
\mu _{\mathbf k}^s(\eta )\simeq A_1^s(k){\rm e}^{-ik\eta }
+A_2^s(k){\rm e}^{ik\eta }\, .
\end{equation}
where $A_1^s(k)$ and $A_2^s(k)$ are two constants that are fixed by
the choice of the initial conditions. Usually, one requires that, on
sub-Hubble scales
\begin{equation}
\mu _{\mathbf k}^s(\eta )=-\frac{4\sqrt{\pi }\ell _{_{\rm Pl}}}{\sqrt{2k}}
{\rm e}^{-ik\left(\eta -\eta _{\rm i}\right)}\, .
\end{equation}
This prescription completely fixes the coefficients $A_1^s(k)$ and
$A_2^s(k)$ which read
\begin{equation}
\label{cidiv}
A_1^s(k)=-\frac{4\sqrt{\pi }\ell _{_{\rm Pl}}}{\sqrt{2k}} {\rm
e}^{ik\eta _{\rm i}}\, , \quad A_2^s(k)=0\, .
\end{equation}
In the above equation, $\ell _{_{\rm Pl}}$ is the Planck length and
$\eta _{\rm i}$ is some initial time at the beginning of
inflation. The knowledge of this time is not important since it will
drop out from the final result.

\par

With the potential given by Eq.~(\ref{approxpotdiv}), the equation of
motion can be solved exactly. Indeed, if we define $\tau \equiv
16i(1+x)/\Theta $ then the equation of motion takes the form
\begin{equation}
\frac{{\rm d}^2 \mu _{\mathbf k}^{_{\rm L}}}{{\rm d}\tau ^2}
+\left[-\frac14+\frac{i\Theta }{16 \tau }+
\frac{1}{4\tau ^2}\right]\mu _{\mathbf k}^{_{\rm L}}=0\, .
\end{equation} 
This is the well-known Whittaker equation, see Eq.~(9.220.1) of
Ref.~\cite{Grad}. The corresponding solution, correctly normalized,
see Eqs.~(\ref{cidiv}), reads
\begin{equation}
\mu _{\mathbf k}^{_{\rm L}}=-\frac{4\sqrt{\pi }\ell _{_{\rm Pl}}}{\sqrt{2k}}
{\rm e}^{ik\eta _{_{\rm i}}}{\rm e}^{-\pi \Theta /32}
{\rm W}_{i\Theta/16, 0}\left[\frac{16i(1+x)}{\Theta }\right]\, ,
\end{equation}
where ${\rm W}_{\kappa ,\mu }(z)$ is the Whittaker function.

\par

Let us now study how the mode function behaves when $x\rightarrow
-1$. The Whittaker function can be expressed in terms of the confluent
hypergeometric function, see Eq.~(13.1.33) of Ref.~\cite{AS}. One
obtains
\begin{eqnarray}
\mu _{\mathbf k}^{_{\rm L}} &=& -\frac{4\sqrt{\pi }
\ell _{_{\rm Pl}}}{\sqrt{2k}}
{\rm e}^{ik\eta _{_{\rm i}}}{\rm e}^{-\pi \Theta /32}
{\rm e}^{-8i(1+x)/\Theta }\sqrt{\frac{16i}{\Theta }(1+x)}
\nonumber \\
& & \times 
U\left[\frac12-i\frac{\Theta }{16},1,\frac{16i}{\Theta }(1+x)\right]\, ,
\end{eqnarray}
where $U(a,b,z)$ is the above-mentioned confluent hypergeometric
function. Using Eq.~(13.5.9) of Ref.~\cite{AS} which says that, when
$z\rightarrow 0$, $U(a,b,z)\rightarrow -[\ln z +\Psi(a)]/\Gamma (a)$,
where $\Gamma (z)$ is the Euler's integral of the second kind and
where $\Psi(z)\equiv {\rm d}\ln \Gamma (z)/{\rm d}z$, see
Ref.~\cite{Grad}, one deduces that
\begin{equation}
\mu _{\mathbf k}^{_{\rm L}} \rightarrow _{x\rightarrow -1}
\sqrt{1+x}\times \ln (1+x)\, .
\end{equation}
But what really matters is not the intermediate variable $\mu
_{\mathbf k}^{_{\rm L}}$ but in fact the amplitude of the
gravitational waves itself given by $h_{\mathbf k}^{_{\rm L}}\equiv
\mu _{\mathbf k}^{_{\rm L}}/z_{_{\rm L}}(\eta )$, see
Eq.~(\ref{defz}). Since $z_{_{\rm L}}\propto \sqrt{1+x}$, one obtains
\begin{equation}
h_{\mathbf k}^{_{\rm L}} \rightarrow _{x\rightarrow -1}
\frac{1}{a(\eta )}\ln (x+1)\, .
\end{equation}
The conclusion is that the amplitude of the mode $({\mathbf k},{\rm
L})$ blows up at the time corresponding to $x=-1$, that is to say at
the time $\eta _{\rm div}(k)$ defined by
\begin{equation}
\eta _{\rm div}(k)\equiv -\frac{8}{k\Theta }\, .
\end{equation}
At this point the linear theory of cosmological perturbations breaks
down and becomes non linear.

\par

An important feature of $\eta _{\rm div}$ is that it is scale
dependent. This means that the physical wavelength of the Fourier
modes $\lambda =(2\pi /k)a(\eta )$, at time $\eta =\eta _{\rm div}$,
are all equal to the same physical length. Explicitly, one has
\begin{equation}
\label{lengththeta}
\frac{\lambda (\eta _{\rm div})}{\ell _{_{\rm Pl}}}
=\frac{\pi }{4}\left(\frac{H_{_{\rm inf}}}{\mP}\right)^{-1}\Theta \, .
\end{equation}
Somehow, this is reminiscent of one of the possible formulations of
the trans-Planckian problem of inflation~\cite{tpl1,tpl2} where it is
postulated that a mode of comoving wavenumber $k$ is ``created'' when
its physical wavelength equals a given new fundamental scale in the
theory (the idea being to test the robustness of the inflationary
predictions to short distance modifications of the theory; therefore,
it is typical in this context to consider that the new scale is the
Planck length, see Refs.~\cite{tpl1} for more details). It is then
easy to show that the ``time of creation'' is inversely proportional
to $k$ as it is the case for $\eta _{\rm div}$, see in particular the
fifth paper in Ref.~\cite{tpl2}. As a consequence, we see from
Eq.~(\ref{lengththeta}) that $\Theta $ defines in fact a new scale the
value of which depends on the inflation scale but also on the string
scale since we will see that the string scale is hidden into the
number ${\cal N}$ which participates to the definition of $\Theta $,
see Eq.~(\ref{deftheta}). A possible way out to the question of the
divergence would be to push the problem in the trans-Planckian
regime. From the above equation, this means that the parameter $\Theta
$ should satisfy
\begin{equation}
\Theta \lsim \frac{H_{_{\rm inf}}}{\mP}\, .
\end{equation}
Therefore, this boils down to a quite stringent constraint on $\Theta
$, typically $\Theta \lsim 10^{-5}$. Unfortunately, we will see in the
following that, for such small values of $\Theta $, the modifications
on large scales are not observable. If one wants to consider larger
values of $\Theta $, it seems that much more refined (\ie non-linear)
calculations are necessary. This calculation are obviously beyond the
scope of the present paper which is just exploratory. 

\par

Let us conclude this subsection by stressing the fact that, a priori,
trans-Planckian effects do not play a deep role in the Chern-Simons
theory under considerations in this work. Here, it is merely a
technical trick which allows to avoid the non-linear regime and to
adopt the common assumption that the Fourier modes emerge from the
trans-Planckian region in the vacuum state. But clearly, if $\Theta
>10^{-5}$, then the non-linear calculation is in principle feasible
without any trans-Planckian considerations.

\subsection{Solutions to the Mode Equation on Very Large Scales}

Let us now study what happens on very large scales, \ie in the limit
where $x$ vanishes. In the situation, the effective potentials can be
very well approximated by the following equations
\begin{eqnarray}
\label{approxpot}
f_{s}(x) &\simeq & \frac{2+3\epsilon }{x^2}+\frac{\lambda
^s}{x}-\frac14\, ,
\end{eqnarray}
where we remind that $\lambda ^{\rm R}=+1$ and $\lambda ^{\rm L}=-1$.
Birefringence enters this equation via the term proportional to $1/x$
which changes its sign according to the polarization state under
considerations. The term proportional to $1/x^2$ is the standard
slow-roll term. The corresponding equation of motion takes the form
\begin{equation}
\frac{{\rm d}^2\mu _{\mathbf k}^s}{{\rm d}x ^2}
+\left[\frac{64}{\Theta ^2}+\frac14-\frac{\lambda ^s}{x}
-\frac{2+3\epsilon }{x^2}\right]\mu _{\mathbf k}^s=0\, .
\end{equation}
Once again, we have to deal with a Whittaker equation. In fact this
equation (and the corresponding power spectrum) has been studied in
details in Ref.~\cite{MS}, see Eq.~(8), and the corresponding power
spectrum has been derived in that reference, see Eq.~(15). Therefore,
in the present paper we can use the results obtained in Ref.~\cite{MS}
and follow the procedure utilized in that reference. Let us introduce
the new definitions
\begin{equation}
y \equiv i \sqrt{1+\frac{256}{\Theta ^2}}x\, ,\quad \kappa \equiv
\frac{i\lambda ^s}{\sqrt{1+256/\Theta ^2}}, \quad \xi \equiv
\frac32+\epsilon \,.
\end{equation}
In the following, we will consider that $256/\Theta ^2\gg 1$ (see the
discussion at the end of the previous subsection) and, as a
consequence, will simply approximate $\sqrt{1+256/\Theta ^2}$ by
$16/\Theta $. With the new definitions taken into account, the
equation of motion takes the form
\begin{equation}
\frac{{\rm d}^2\mu _{\mathbf k}^s}{{\rm d}y^2}
+\left[-\frac14+\frac{\kappa}{y }+\left(\frac14-\xi ^2\right)
\frac{1}{y^2}\right]\mu _{\mathbf k}^s=0\, ,
\end{equation} 
which is again the Whittaker equation. The situation is exactly
similar to the one studied around Eq.~(12) of Ref.~\cite{MS}. The
exact general solution to this equation is given in terms of Whittaker
functions
\begin{equation}
\mu _{\mathbf k}^s(\eta )=C_1^s(k)W_{\kappa, \xi}\left(y\right)
+C_2^s(k)W_{-\kappa, \xi}\left(-y\right)\, ,
\end{equation}
where $C_1^s(k)$ and $C_2^s(k)$ are two constants fixed by the choice
of the initial conditions. 

\par

As discussed in the preceding subsection, we will fix the initial
conditions in the region $-1<x\ll 0$ which is free of divergences. In
this regime, the only natural choice that we have is to postulate a
plane wave. This is equivalent to postulating that the non-linear
phenomena occurring around the divergence of the effective potential,
provided they happen in the trans-Planckian region, will not affect
the standard choice of the initial conditions in the region
$x>-1$. Somehow, this is the same assumption that is made in the
standard inflationary scenario. Indeed, despite the fact that the
modes of astrophysical interest today originate from the
trans-Planckian region~\cite{tpl1}, the vacuum is assumed to be the
correct initial state. Let us stress, however, that a possible
weakness of the above comparison is that, in the case of the
trans-Planckian problem of inflation~\cite{tpl1,tpl2}, one can show
that the final result can be robust to changes in the short distance
physics~\cite{tpl1,tpl2} (under some conditions, \ie adiabatic
evolution of the Fourier modes). In the present context, however, it
is more difficult to imagine that the non-linearities will not affect
the initial conditions. On the other hand, in the absence of a
second-order calculations and as a first approach to the problem, this
seems to be quite reasonable. As shown in Ref.~\cite{MS}, see
Eqs.~(14), this choice amounts to
\begin{eqnarray}
C_1^s(k) &=& -\frac{4\sqrt{\pi }\ell _{_{\rm Pl}}}{\sqrt{2k}} {\rm
e}^{iq\eta _{\rm i}}\exp\left(-\frac{\lambda ^s\pi
\Theta}{32}\right)\, , 
\\
C_2^s(k) &=& 0\, ,
\end{eqnarray}
where we have used the fact that, see Eq.~(9.227) of Ref.~\cite{Grad},
$\lim _{\vert y \vert \rightarrow +\infty }W_{\kappa, \xi}(y)={\rm
e}^{-y/2} y^{\kappa }$. The sign of the argument in the exponential
depends on the polarization state considered, as expected. We conclude
that the solution to the mode equation on very large scales is now
known explicitly.

\subsection{The Power Spectrum}

Usually, the power spectrum is given by the two-point correlation
function calculated in the vacuum state. Another way to calculate the
same quantity is to view it as a classical spatial average. Since a
fully consistent quantum formulation of the present theory is not yet
available, we adopt the second point of view. Therefore, the two-point
correlation function can be written as
\begin{equation}
\langle h_{ij}\left(\eta ,{\mathbf x}\right)
h^{ij}\left(\eta ,{\mathbf x}\right)\rangle 
=\frac{1}{V}\int {\rm d}{\mathbf x}\, h_{ij}\left(\eta ,{\mathbf x}\right)
h^{ij}\left(\eta ,{\mathbf x}\right)\, ,
\end{equation}
with $V=\int {\rm d}{\mathbf x}$ is the total volume. Using the
properties of the polarization tensor, straightforward calculations
show that
\begin{equation}
\langle h_{ij}\left(\eta ,{\mathbf x}\right)
h^{ij}\left(\eta ,{\mathbf x}\right)\rangle 
=\frac{1}{\pi ^2}\sum _{s={\rm L},{\rm R}}\int _0^{+\infty }
\frac{{\rm d}k}{k}k^3\left\vert h_{\mathbf k}^s\right \vert ^2\, ,
\end{equation}
from which we deduce the power spectrum
\begin{equation}
\label{spec}
k^3P_h^s(k)=\frac{k^3}{\pi ^2}\left\vert \frac{\mu _{\mathbf
k}^s}{a(\eta )\sqrt{1-\lambda ^skf'/a^2}}\right \vert ^2\, .
\end{equation}
Let us notice that, usually, the power spectrum is proportional to the
factor $2k^3/\pi ^2$. Here we don't have the factor $2$ because we
consider the two states of polarization separately (\ie usually, these
two states are summed and produce the factor $2$).

\par

A priori, using the solution obtained in the previous subsection, we
can calculate the spectrum exactly in terms of the Whittaker
function. But only the spectrum on large scales is needed and in this
regime one has (for details, see Ref.~\cite{MS})
\begin{equation}
k^3P_h^s=\frac{16}{\pi}\frac{\ell _{_{\rm Pl}}^2}{\ell _0^2}
\frac{k^{-2\epsilon }}{2^{2\xi }} \frac{\Gamma ^2\left(2\xi
\right)}{\vert \Gamma \left(1/2+\xi -i\lambda ^s\Theta /16\right)\vert
^2}{\rm e}^{-\lambda ^s\pi \Theta /16}\, .
\end{equation}
Let us notice that we have neglected the factor $(1-\lambda
^skf'/a^2)^{-1/2}$ because $kf'/a^2$ is proportional to $k\eta $ and
hence negligible on large scales. The above expression is similar to
Eq.~(15) of Ref.~\cite{MS}. At this stage, the only thing which
remains to be done is to expand the above expression at first order in
the slow-roll parameter. After lengthy but straightforward
calculations, one obtains the following result
\begin{widetext}
\begin{eqnarray}
k^3P_h^s(k) &=& \frac{16H_{_{\rm inf}}^2}{\pi m_{_{\rm Pl}}^2}
\frac12 {\cal A}^s\left(\Theta \right)
\left[1-2\left(C+1\right)\epsilon -2\epsilon \ln
\frac{k}{k_*}-\epsilon {\cal B}(\Theta )\right] \, ,
\end{eqnarray}
\end{widetext}
with, 
\begin{eqnarray}
{\cal A}^s(\Theta ) &\equiv & \frac{16}{\pi \Theta
}\left(1+\frac{\Theta ^2}{256}\right)^{-1} \sinh \left(\frac{\pi
\Theta }{16}\right)\nonumber 
\\
& & \times 
\exp\left(-\lambda ^s\frac{\pi \Theta}
{16}\right)\, ,
\\ 
{\cal B}(\Theta ) &\equiv & \Psi \left(2-i\frac{\Theta}{16}\right)+\Psi
\left(2+i\frac{\Theta}{16}\right)-2\Psi(2)\, .
\end{eqnarray}
At this point some remarks are in order. As required one can check
that, when $\Theta =0$, the standard inflationary result is recovered.
This is the case because ${\cal A}^s(0)=1$ and ${\cal B}(0)=0$. As
already mentioned, a factor $1/2$ is left because the (now identical)
contribution from the two states of polarization should be added. The
function ${\cal A}^s$ describes the dominant modification in the
amplitude of the power spectrum (the contribution originating from the
function ${\cal B}$ is clearly sub-dominant since it is proportional
to the slow-roll parameter $\epsilon $). For small values of $\Theta $
we have
\begin{eqnarray}
{\cal A}^{\rm R} &=& 1- \frac{\pi }{16}\Theta + \left(\frac{\pi
^2}{384}-\frac{1}{256}\right)\Theta ^2 +{\cal O}\left(\Theta
^3\right)\, ,
\\
{\cal A}^{\rm L} &=& 1+ \frac{\pi }{16}\Theta + \left(\frac{\pi
^2}{384}-\frac{1}{256}\right)\Theta ^2 +{\cal O}\left(\Theta
^3\right)\, ,
\end{eqnarray}
where $\pi/16\simeq 0.2$ and $(\pi ^2/384-1/256)\simeq
0.022$. Therefore, the amplitude of the right polarization state is
reduced while the one of the left polarization state is
enhanced. However, for small values of $\Theta $, the effect is
clearly not very important.

\par

Another conclusion that can be obtained from the above spectrum is
that, at leading order in the slow roll parameter, the spectral index
remains unmodified. Indeed, one has $ n_{_{\rm T}}^s={\rm d}\ln
\left(k^3P_h^s\right)/{\rm d}\ln k=-2\epsilon $ for each polarization
state.

\par

Finally, let us now compute how the ratio $T/S$ is modified. The
scalar power spectrum is not modified (see also Ref.~\cite{hwang}) and
reads~\cite{MS2}
\begin{equation}
k^3P_{\zeta }=\frac{H^2_{_{\rm inf}}}{\pi m_{_{\rm Pl}}^2\epsilon }
\left[1-2\epsilon -2C(2\epsilon -\delta )-2(2\epsilon -\delta ) \ln
\frac{k}{k_*}\right]\, .
\end{equation}
Therefore, we conclude that the consistency check of inflation, at
leading order in the slow-roll parameters, can now be written as
\begin{eqnarray}
\frac{T}{S} &\equiv & \frac{1}{\left(k^3P_{\zeta }\right)}\left(\sum
_{s={\rm L},{\rm R}} k^3P_h^s\right)\Biggl\vert _{k=k_*} 
\\ 
&=&
16\epsilon \times \frac12 \left[{\cal A}^{\rm L}\left(\Theta
\right)+{\cal A}^{\rm R} \left(\Theta \right)\right]
\\
&\simeq & 16\epsilon \times \left[1+\left(\frac{\pi
^2}{384}-\frac{1}{256}\right)\Theta ^2\right]\, .
\end{eqnarray}
Unfortunately, the linear corrections in $\Theta $ cancels out and we
are left with a correction which is quadratic in $\Theta $. Another
way to express the above result is to calculate the ratio of $T/S$
with the Chern-Simon modification taken into account to $T/S$ obtained
in the standard case. One gets
\begin{equation}
\label{cccheck}
\frac{\left(T/S\right)_{\Theta \neq 0}}{\left(T/S\right)_{\Theta =0}}
\simeq 1+0.022 \times \Theta ^2\, .
\end{equation}
It is clear from this expression that the modification is not
observable at all since we have seen before that, typically, $\Theta
\lsim 10^{-5}$ in order for the calculations presented here to be
consistent (\ie for the divergence of the effective potential to be in
the trans-Planckian region).

\section{Discussion and Conclusions}

We have evaluated the super-Hubble power spectrum and the tensor to
scalar ratio for birefringent gravitational waves produced during
inflation. The power spectrum exhibits two interesting regimes, linear
and non-linear.  The non-linear regime occurs when $k\eta \sim \Theta
^{-1}$ because the effective potential controlling the evolution of
the linear perturbations blows up. At this point, the linear theory of
cosmological perturbations that we used is no longer valid. This
divergence occurs for all modes (\ie for all comoving wavenumber $k$)
but at different times.

\par

In this present investigation we only considered the linear regime
since at the present moment we were not able to perform a rigorous
analysis of the non-linear phenonema. We found corrections which
survive to second order in $\Theta$.  Therefore, in this regime the
tensor to scalar ratio gets corrected by $\Theta$ but this effect is
very small.

\par 

If $\Theta \lsim 10^{-5}$, one can push the non-linear regime (\ie the
divergence in the effective potential) into the trans-Planckian region
where, anyway, other effects (for instance, non perturbative stringy
effects) are likely to become important. Somehow, this corresponds to
the standard situation where the evolution from the Planck scale to
the super-horizon scales is under control and where the perturbations
are assumed to emerge from the trans-Planckian regime in the vacuum
state, thus ignoring the modifications of the initial conditions that
the trans-Planckian physics could cause (in the very same way that we
have ignored the effect of the divergence in the potential, provided
it is in the trans-Planckian region). However, it is important to keep
in mind that this is mostly a technical trick which allows us to work
with the linear theory. At a deeper level, the trans-Planckian effects
are not expected to play a more important role than in the standard
situation. In particular, if the divergence is not in the
trans-Planckian regime, only the non-linear theory of cosmological
perturbations is necessary in order to calculate the modified $T/S$
irrespectively of any trans-Planckian effects.

\par

It is interesting to note that the linear regime (where $\Theta \sim
10^{-5}$) is compatible with the Stringy Embedding of inflationary
Baryogenesis (SEB)~\cite{Gates}. In this context, the value of
$\Theta$ enhances and gives the resonant frequency associated with the
observed baryon asymmetry. As already mentioned before, this value is
completely fixed by the string scale and coupling in a model
independent fashion. Explicitly, the value of the number ${\cal N}$
which appears in the definition of $\Theta $, see
Eq.~(\ref{deftheta}), is given by
\begin{equation} 
{\cal N}= \pi ^2\sqrt{\frac{g_{\rm s}}{2}} \left(\frac{M_{_{\rm
Pl}}}{M_{_{\rm 10}}}\right)^2\, ,
\end{equation}
where $M_{_{\rm 10}}$ is the ten-dimensional fundamental scale and
$g_{\rm s}$ is the string coupling. Therefore, we established a direct
link between stringy quantities and CMB anisotropies. Explicitly,
Eq.~(\ref{cccheck}) can be re-written as
\begin{equation}
\frac{\left(T/S\right)_{\Theta \neq 0}}{\left(T/S\right)_{\Theta =0}}
\simeq 1+\frac{0.022}{4}\left(\frac{H_{_{\rm inf}}}{M_{\rm
10}}\right)^4 g_{\rm s}\epsilon \, .
\end{equation}
In a recent paper the authors of Ref.~\cite{Gates} found that for
reasonable values of string coupling (\ie weak) and the string scale
set to $10^{16}{\rm GeV}$, both $\Theta $ can be as small as $10^{-2}$
and the observed baryon asymmetry can be generated. Of course the
stringy embedding admits much larger values of $\Theta$ putting our
analysis into the non-linear regime.

\par

If $\Theta \gsim 10^{-5}$, a non-linear calculation is mandatory and
one can hope to obtain a significative modification of the ratio
$T/S$, maybe observable by future high accuracy CMB
experiments. Clearly, the present situation is not very satisfactory
since the regime for which sizable effects are expected turns out to
be very complicated from the technical point of view. Furthermore, we
suspect that the scale associated to the divergence of the effective
potential corresponds to a resonant production of lepton number. We
wish to report on this issue in a future paper.

\vspace{0.5cm} 
\centerline{\bf Acknowledgments}
\vspace{0.2cm}

We wish to thank R.~Brandenberger, M.~Peskin and P.~Peter for many
enlightening comments and discussions.


\begin{thebibliography}{100}

\bibitem{APS} S.~Alexander, M.~Peskin and M.~Sheikh-Jabbari, {\tt
hep-th/0403069}

\bibitem{Gates}
S.~H.~S.~Alexander and S.~J.~J.~Gates, {\tt hep-th/0409014}.

\bibitem{Lue} 
A.~Lue, L.~M.~Wang and M.~Kamionkowski, Phys.~Rev.~Lett.~{\bf 83},
1506 (1999), {\tt astro-ph/9812088}.

\bibitem{Pogosian}
L.~Pogosian, T.~Vachaspati and S.~Winitzki, New~Astron.~Rev.~{\bf 47},
859 (2003), {\tt astro-ph/0210039}.

\bibitem{Balaji}
K.~R.~S.~Balaji, R.~H.~Brandenberger and D.~A.~Easson, JCAP~{\bf
0312}, 008 (2003) {\tt hep-ph/0310368}.

\bibitem{hwang} K.~Choi, J.~Hwang and K.~Hwang, Phys.~Rev.~D {\bf 61},
084026 (2000), {\tt hep-ph/9907244}.

\bibitem{MS2} J.~Martin and D.~J.~Schwarz, Phys.~Rev.~D {\bf 62},
103520 (2000), {\tt astro-ph/9911225}; J.~Martin and D.~J.~Schwarz,
Astrophys.~J. {\bf 543}, L99 (2000), {\tt astro-ph/0006392}.

\bibitem{MS} J.~Martin and D.~J.~Schwarz, Phys.~Lett.~B {\bf 500}, 1
(2001), {\tt astro-ph/0005542}.

\bibitem{Grad} I.~S.~Gradshteyn and I.~M.~Ryshik, {\sl Table of
Integrals, Series and Products}, (Academic Press, New York, 1980).

\bibitem{AS} M.~Abramowitz and I.~A.~Stegun, {\sl Handbook of
Mathematical Functions}, (Dover publications, New York, 1964).

\bibitem{tpl1} J.~Martin and R.~H.~Brandenberger, Phys.~Rev.~D {\bf
63}, 123501 (2001), {\tt hep-th/0005209}; R.~H.~Brandenberger and
J.~Martin, Mod.~Phys.~Lett. {\bf A16}, 999 (2001), {\tt
astro-ph/0005432}.

\bibitem{tpl2} M.~Lemoine, M.~Lubo, J.~Martin and J.~P.~Uzan,
Phys.~Rev.~D {\bf 65}, 023510 (2002), {\tt hep-th/0109128};
J.~C.~Niemeyer, Phys.~Rev.~D {\bf 63}, 123502 (2001), {\tt
astro-ph/0005533}; A.~Kempf, Phys.~Rev.~D {\bf 63}, 083514 (2001),
{\tt astro-ph/0009209}; R.~Easther, B.~R.~Greene, W.~H.~Kinney and
G.~Shiu, Phys.~Rev.~D {\bf 64}, 103502 (2001), {\tt hep-th/0104102};
J.~Martin and R.~H.~Brandenberger, Phys.~Rev.~D {\bf 68}, 063513
(2003), {\tt hep-th/0305161}.  J.~Martin and C.~Ringeval, Phys.~Rev.~D
{\bf 69}, 083515 (2004), {\tt astro-ph/0310382}; J.~Martin and
C.~Ringeval, Phys.~Rev.~D {\bf 69}, {\tt astro-ph/0402609}; J.~Martin
and C.~Ringeval, {\tt astro-ph/0405249}.








\end{thebibliography}
\end{document}